\documentclass{article}

 \usepackage{arxiv}
 
 \usepackage[utf8]{inputenc} % allow utf-8 input
 \usepackage[T1]{fontenc}    % use 8-bit T1 fonts
 \usepackage{hyperref}       % hyperlinks
 \usepackage{url}            % simple URL typesetting
 \usepackage{booktabs}       % professional-quality tables
 \usepackage{amsfonts}       % blackboard math symbols
 \usepackage{nicefrac}       % compact symbols for 1/2, etc.
 \usepackage{microtype}      % microtypography
 \usepackage{lipsum}     % Can be removed after putting your text content
 \usepackage{graphicx}
 \usepackage[numbers]{natbib}
 \usepackage{doi}
 \usepackage{amssymb}
 \usepackage{amsmath} % for \boldsymbol macro
 \usepackage{bm} 
 \usepackage{moreverb}
\newcommand\vWFc{vWF\textsubscript{c}~}
\newcommand\vWFs{vWF\textsubscript{s}~}
\newcommand\vWFsCrit{$\mathrm{vWF_{s}^{crit}}$~}
\newcommand\bs{\boldsymbol{s}}
\newcommand\bu{\boldsymbol{u}}
\newcommand\by{\boldsymbol{y}}
\newcommand\bX{\boldsymbol{X}}
\newcommand\bY{\boldsymbol{Y}}
\newcommand\argmin{argmin}

\title{Uncertainty quantification of a thrombosis model \\ considering the clotting assay PFA-100\textregistered}

\author{
  Rodrigo M\'endez Rojano \\
  Meinig School of Biomedical Engineering \\
  Cornell University\\
  Ithaca, NY 14850 \\
  \texttt{rm2235@cornell.edu} \\
   \And
  Mansur Zhussupbekov \\
  Meinig School of Biomedical Engineering \\
  Cornell University\\
  Ithaca, NY 14850 \\
  \texttt{mz332@cornell.edu} \\
  \And 
  James F. Antaki\\
  Meinig School of Biomedical Engineering \\
  Cornell University\\
  Ithaca, NY 14850 \\
  \texttt{antaki@cornell.edu} \\
  \And
  Didier Lucor \\
  Laboratoire Interdisciplinaire des Sciences du Num\'erique \\
  CNRS, Universit\'e Paris-Saclay \\
  Orsay, France \\
  \texttt{didier.lucor@lisn.upsaclay.fr} \\
}

\begin{document}

\maketitle

\begin{abstract}
Mathematical models of thrombosis are currently used to study clinical scenarios of pathological thrombus formation. Most of these models involve inherent uncertainties that must be assessed to increase the confidence in model predictions and identify avenues of improvement for both thrombosis modeling and anti-platelet therapies. In this work, an uncertainty quantification analysis of a multi-constituent thrombosis model is performed considering a common assay for platelet function (PFA-100\textregistered). The analysis is performed using a polynomial chaos expansion as a parametric surrogate for the thrombosis model. The polynomial approximation is validated and used to perform a global sensitivity analysis via computation of Sobol' coefficients. Six out of fifteen parameters were found to be influential in the simulation variability considering only individual effects. Nonetheless, parameter interactions are highlighted when considering the total Sobol' indices. In addition to the sensitivity analysis, the surrogate model was used to compute the PFA-100\textregistered ~closure times of 300,000 virtual cases that align well with clinical data. The current methodology could be used including common anti-platelet therapies to identify scenarios that preserve the hematological balance. 
\end{abstract}


\section{Introduction}\label{sec:Intro}
Thrombosis, which is defined as excessive formation of blood clot or thrombus, is a common pathology in several cardiovascular diseases \cite{Nagareddy2013}, and blood-wetted medical devices \cite{DalSasso2019,Jaffer2019,Labarrere2020}. Thrombus formation is characterized by an intertwined process of platelet activity and coagulation. In this hemostatic process, platelet activation and aggregation leads to formation of a platelet plug that is mechanically stabilized by a fibrin net formed by coagulation reactions \cite{Colman2006}. In hemodynamic conditions that promote high shear stresses, such as stenotic vessels or prosthetic heart valves or blood pumps, von Willebrand Factor (vWF) plays a central role in thrombosis. vWF is a blood glycoprotein that has a collapsed globular conformation in its natural state and unfolds in response to extensional flow and high-shear flow conditions. Once unfolded, vWF plays an important role in the platelet adhesion and aggregation process enabling clot formation.

Multi-scale computational models of thrombosis have been developed in the past decades to understand and predict the thrombus formation dynamics in academic and clinical configurations (\cite{Leiderman2011a,Menichini2016,Yazdani:2017,Xu2008,Kim2013,Xu2017,Yazdani2018}). These models span several degrees of complexity, from simple flow characteristic indices to complicated, coupled biochemical interactions. Mechanistic models, for example, may contain equations that describe shear dependent platelet activity including adhesion, aggregation, and activation \cite{Sorensen1999,Storti2014,Danes2019}; hemodynamics that regulate the transport of biochemical species \cite{Storti2014b,Fogelson2015}; and/or coagulation reactions that lead to the formation of fibrin \cite{Anand2006b,Yesudasan2019,Hansen2019}.
Increasing modeling complexity improves the fidelity and versatility of the model, but the computational cost of the simulation escalates.
Also, increasing detail introduces additional uncertainties that can have a great impact on its accuracy \cite{Belyaev2018}. These uncertainties are related to the fundamental structure of the model \cite{}, numerical discretization, and/or model parameters such as diffusion coefficients, concentrations of biochemical species, fluid viscosity, etc. Previous investigators have introduced several uncertainty quantification (UQ) strategies to evaluate the effect of such uncertainties on model performance in the context of blood flows \cite{Nikishova2018,Sankaran2016,Steinman2018}. UQ analyses in thrombosis models have been performed to identify the sensitivity of input parameters \cite{Danforth2009,MendezRojano2019a,Link2020,Melito2020}. In most sensitivity studies, the global Morris method \cite{Campolongo2007} or the Sobol' indices method \cite{Saltelli2010} have been used. The Morris method is used as a screening tool to identify important parameters in models with a large number of input variables. The Sobol' method produces indices based on variance that quantifies the output uncertainty related to each input parameter. For example, Melito et al. \cite{Melito2020} performed a variance based sensitivity analysis on the thrombosis model of Menichini et al. \cite{Menichini2016} considering thrombosis in a backward facing step. Their results suggest that only four of the nine parameters included in their thrombosis model had significant impact in thrombus formation implying that the model could be made sparser to optimize computational resources. Link et al. \cite{Link2020b} performed a sensitivity analysis to identify important coagulation factors in the thrombin generation profile among hemophilia A patients. Their work allowed the identification of coagulation factor V as an important modifier of thrombin formation among patients with hemophilia A, paving the way to pharmacological interventions to treat these patients. In addition, as demonstrated by Link et al. \cite{Link2020b}  the synthetic databases built to perform sensitivity analysis can serve as hypothesis generation tool for \textit{in-vitro} experiments or, in the long run, patient treatment.

In this work, a forward uncertainty propagation study is conducted using a well-established but computationally costly multi-constituent thrombosis model introduced by Wu et al. \cite{Wu2017}. A widely used platelet function assay PFA-100\textregistered~ (Siemens, Erlangen, Germany) was selected as the study case. To understand the impact of uncertain input parameters a variance based global sensitivity analysis was conducted monitoring Sobol' sensitivity coefficients taking advantage of a polynomial chaos approximation as a surrogate of the baseline thrombosis model. The surrogate model is also used in this study to assess its predictive ability of the PFA-100\textregistered ~closure time distribution in three scenarios: normal platelet count, thrombocytopenia, and vWF deficiency. The predicted closure times distribution align well with clinical data.

\section{Methods \label{sec:Methods}}
\subsection{Baseline Thrombosis Model}
\begin{figure}[h!]
\centering
\includegraphics[width=0.8\textwidth]{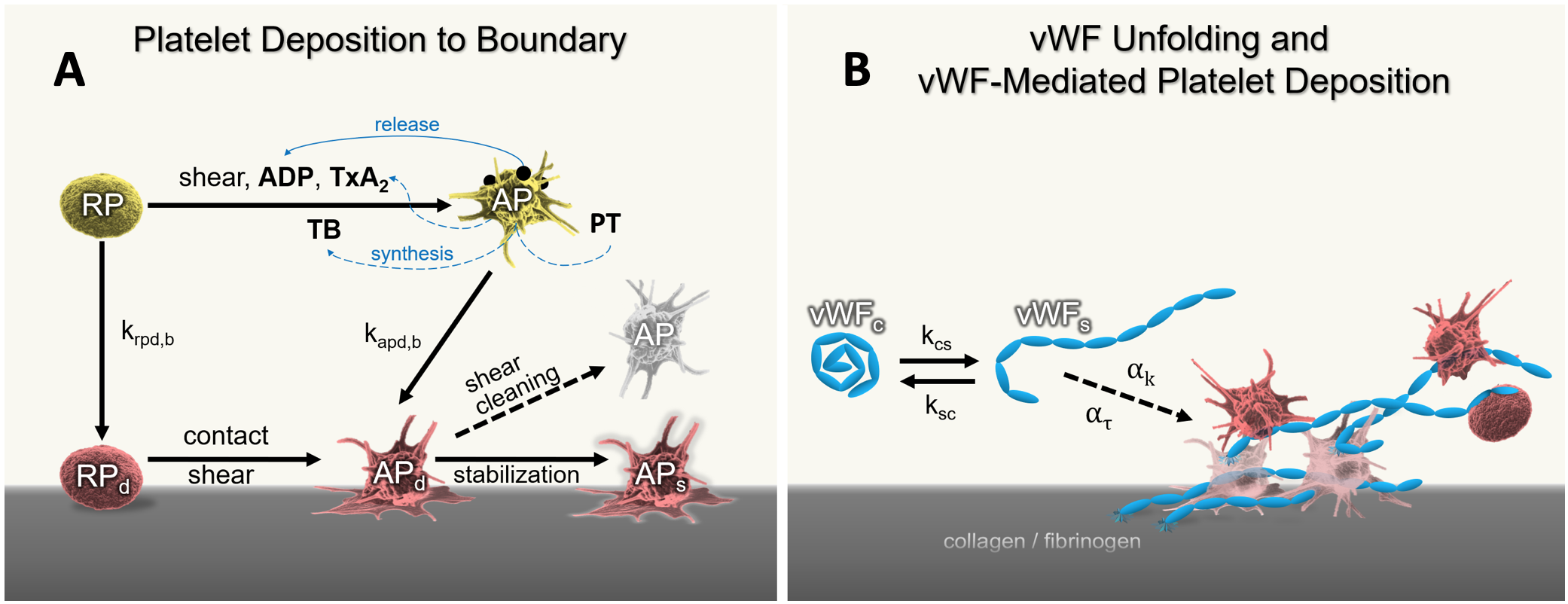}
\caption{\label{Fig:ThrombosisDiagram} A) Diagram of platelet activation and deposition in the thrombosis model. $k_{rpd,b}$ and $k_{apd,b}$ are the rates at which resting and activated platelets deposit to the surface. Resting platelets can be activated by mechanical shear, or the combination of agonists: ADP, TxA$_2$ and thrombin. Once deposited, AP$_d$ can stabilize or embolize due to flow shearing forces. B) Schematic depiction of vWF unfolding and vWF-mediated platelet deposition and aggregation. The presence of stretched \vWFs amplifies the deposition rate of free-flowing platelets by $\alpha_{k}$ and increases the resistance of deposited platelets to shear cleaning by $\alpha_{\tau}$.}
\end{figure}

The thrombosis model used in the current work is a variation of the approach introduced by Wu et al. \cite{Wu2017} with the addition of vWF activity. Figure~\ref{Fig:ThrombosisDiagram} A shows how platelets, the central player of the model, deposit to artificial surfaces. Figure~\ref{Fig:ThrombosisDiagram} B illustrates vWF unfolding due to high shear rates increasing platelet deposition. Thrombus growth is quantified through platelet deposition over artificial surfaces and subsequent platelet aggregation. Several biochemical species are considered in the model: 1) \textbf{AP} Activated platelets, 2) \textbf{RP} Resting Platelets, 3) \textbf{AP$_d$} Deposited activated platelets, 4) \textbf{RP$_d$} Deposited Resting Platelets 5) \textbf{ADP} Adenosine Diphosphate, 6) \textbf{TxA$_2$} Thromboxane A2, 7)\textbf{TB} Thrombin , 8) \textbf{PT} Prothrombin, 9) \textbf{AT} Anti-Thrombin, 10) \textbf{\vWFc} vWF Coiled, and 11) \textbf{\vWFs} vWF stretched.

The spatial and temporal dynamics of biochemical species are quantified through convection-diffusion-reaction equations of the form:

\begin{equation}\label{eq:CDR}
\frac{\partial c_i}{\partial t} =  \nabla \cdot (D_i \nabla c_i) - \boldsymbol{v}_f  \cdot \nabla c_i + r_i ,
\end{equation}

where: $c_i $ is the concentration of species $i$, $D_i$ is the diffusion coefficient, $\boldsymbol{v}_f$ is the velocity vector field, and $r_i$ is the reaction source term that accounts for biochemical interactions such as platelet activation by thrombin or TxA$_2$.

\subsubsection{vWF activity}
To account for the role of vWF in high shear stress thrombosis the original model of Wu et al.\cite{Wu2017} was amended to include the enhanced platelet aggregation and adhesion driven by vWF activity. The right-hand side of Figure~\ref{Fig:ThrombosisDiagram} shows the two inter-convertible states of vWF in the model: collapsed \vWFc and stretched \vWFs. In all simulations, only \vWFc is introduced at the inlet of the domain and \vWFs is produced by means of \vWFc unfolding. The presence of stretched \vWFs polymers has a local thrombogenic effect via two mechanisms: it amplifies the deposition rate of free-flowing platelets by scaling it by $\alpha_{k}$ (Eq~(\ref{eq:kamp})) and increases the resistance of deposited platelets to shear cleaning by scaling the shear-cleaning parameters, $\tau_{emb}$ and $\tau_{emb,b}$, by $\alpha_{\tau}$ according to Eq~(\ref{eq:tauAmp})

\begin{equation}
\label{eq:kamp}
	\alpha_{k}=\begin{cases}
		1, & \text{if \vWFs $\leq$ \vWFsCrit}\\
		\frac{\textrm{\vWFs}}{\textrm{\vWFsCrit}}, & \text{if \vWFs $>$ \vWFsCrit}
	  \end{cases}
\end{equation}
 
\begin{equation}
\label{eq:tauAmp}
	\alpha_{\tau}=\begin{cases}
		1, & \text{if \vWFs $\leq$ \vWFsCrit}\\
		\frac{D_e(\dot{\gamma})}{D^{l}_{e}}, & \text{if \vWFs $>$ \vWFsCrit},
	  \end{cases}
\end{equation}

where \vWFsCrit is set as 5\% of the total vWF concentration; $D_e$ is the Morse potential's well depth proposed by Yazdani et al. \cite{Yazdani:2017} that correlates the platelet adhesion force to the local shear rate, $\dot{\gamma}$, according to Eq~(\ref{eq:DeYazdani});

\begin{equation}
\label{eq:DeYazdani}
	D_{e}(\dot{\gamma})=D_{e}^{h}\left[\tanh\left(\frac{\dot{\gamma}-\dot{\gamma}_{vW\!F\!}}{1000}\right)+\frac{D_{e}^{l}}{D_{e}^{h}}+1\right],
\end{equation}

$D^{l}_{e}$ and $D^{h}_{e}$ determine the adhesive forces at low and high shear rates, respectively, and $\dot{\gamma}_{vW\!F\!}$ = 5500 s\textsuperscript{-1} is the critical shear rate for vWF unfolding and marks the transition from low-shear to high-shear regime in this expression. 

\vWFc unfolding occurs via two mechanisms: periodic tumbling in simple shear and strong unfolding in flows with extensional kinematics. The flow classification is based on polymer unfolding criteria by Babcock et al. \cite{Babcock2003} where they define the flow-type $\lambda$ as: 

\begin{equation}
\label{eq:flowType}
	\lambda=\frac{\|\boldsymbol{S}\|-\|\boldsymbol{\varOmega}\|}{\|\boldsymbol{S}\|+\|\boldsymbol{\varOmega}\|}
\end{equation}

where  $\|\boldsymbol{S}\|$ and $\|\boldsymbol{\varOmega}\|$ are the magnitudes of the symmetric and anti-symmetric parts of the velocity gradient tensor, respectively. Thus, in the limiting cases of the flow type parameter, $\lambda=1$ corresponds to purely extensional flow, $\lambda=0$ to simple shear flow, and $\lambda=-1$ to solid-body rotation. vWF will experience strong unfolding in flows with $\lambda\geq\lambda_{crit}$, simple-shear tumbling in $-\lambda_{crit}\leq\lambda<\lambda_{crit}$, and remain collapsed if $\lambda<-\lambda_{crit}$.   
 
In simple shear and near-shear flows, $-\lambda_{crit}\leq\lambda<\lambda_{crit}$, the unfolding rate, $k_{c\text{-}s}$ (collapsed-to-stretched conversion), follows a function proposed by Lippok et al. \cite{Lippok2016} for shear-dependent cleavage rate of vWF. Since this function describes the availability of vWF monomers for enzymatic cleavage, the unfolding rate is assumed to follow the same shape, given by Eq.~(\ref{eq:kcsShear})

\begin{equation}
\label{eq:kcsShear}
	k^{shear}_{c\text{-}s}(\dot{\gamma})=\frac{k^{\prime}}{1+\exp\left(-\frac{\dot{\gamma}-\dot{\gamma}_{vW\!F\!}}{\Delta\dot{\gamma}}\right)}
\end{equation}

\begin{equation}
\label{eq:kscShear}
	k^{shear}_{s\text{-}c}=k^{\prime}
\end{equation}

where $\dot{\gamma}_{vW\!F\!}$ = 5500 s\textsuperscript{-1} is the critical (half-maximum) shear rate for vWF unfolding and $\Delta\dot{\gamma}$ = 1271 s\textsuperscript{-1} denotes the width of the transition. $k^{\prime}$ is the nominal state conversion rate, given by $k^{\prime}=1/t_{vWF}$, where $t_{vWF}$ = 50 ms is the vWF unfolding time. This timescale for vWF unfolding was reported in experiments by Fu et al.\cite{Fu2017a} and Brownian dynamics simulations by Dong et al. \cite{Dong2019,Kania2021} To reflect the tumbling behavior of the vWF in this regime, the stretched-to-collapsed conversion rate, $k_{s\text{-}c}$ is set to the nominal value $k^{\prime}$ (Eq.~(\ref{eq:kscShear})).

In extensionally-dominated flows with $\lambda\geq\lambda_{crit}$, strong unfolding occurs if the modified Weissenberg number, $Wi_{ef\!f\!}$, given by Eq.~(\ref{eq:WiEff}), exceeds the critical value $Wi_{ef\!f\!,crit}$. Then, the unfolding rate $k_{c\text{-}s}^{steady}$ scales with $Wi_{ef\!f\!}$ according to Eq.~(\ref{eq:kcsSteady}), while $k_{s\text{-}c}^{steady}$ is zero. 

If $Wi_{ef\!f\!}$ falls below $Wi_{ef\!f\!,crit}$, the stretched \vWFs exhibits hysteresis and does not collapse back to \vWFc until $Wi_{ef\!f\!}<Wi_{ef\!f\!,hyst}$, as shown in Eq.~(\ref{eq:kscSteady}). 

\begin{equation}
\label{eq:WiEff}
	Wi_{ef\!f\!}=\sqrt{\lambda}\left(\|\boldsymbol{S}\|+\|\boldsymbol{\varOmega}\|\right)\tau_{rel}
\end{equation}

\begin{equation}
\label{eq:kcsSteady}
	k_{c\text{-}s}^{steady}=\begin{cases}
		k^{\prime}\frac{Wi_{ef\!f\!}}{Wi_{ef\!f\!,crit}}, & \text{if $Wi_{ef\!f\!} \geq Wi_{ef\!f\!,crit}$}\\
		k^{shear}_{c\text{-}s}, & \text{if $Wi_{ef\!f\!} < Wi_{ef\!f\!,crit}$}
	  \end{cases}
\end{equation}

\begin{equation}
\label{eq:kscSteady}
	k_{s\text{-}c}^{steady}=\begin{cases}
		0, & \text{if $Wi_{ef\!f\!,hyst} \leq Wi_{ef\!f\!} < Wi_{ef\!f\!,crit}$}.\\
		k^{\prime}, & \text{if $Wi_{ef\!f\!} < Wi_{ef\!f\!,hyst}$}.
	  \end{cases}
\end{equation}

where $\tau_{rel}$ is the polymer relaxation time used to non-dimensionalize the expression.
Here, we adopted the unfolding threshold and the hysteresis value reported by Sing and Alexander-Katz.\cite{Sing2011b} Assuming the monomer diffusion time of $1.02\times10^{-3}$ s for $\tau_{rel}$ in Eq.~(\ref{eq:WiEff}), $Wi_{ef\!f,crit}=0.316$ and $Wi_{ef\!f,hyst}=0.053$. 

If $\lambda<-\lambda_{crit}$, \vWFc remains collapsed and \vWFs reverts to a globular state, so $k_{c\text{-}s}=0$ and $k_{s\text{-}c}=k^{\prime}$.

A detailed validation of the vWF-thrombosis model is provided in Zhussupbekov et al. \cite{Zhussupbekov2021} for multiple academic simple flow configurations and complex in-vitro clotting scenarios. 

\subsubsection{Flow dynamics}
The pressure and velocity fields $p$ and $\boldsymbol{v}_f$ are obtained solving the Navier-Stokes equations considering an incompressible Newtonian fluid: 

\begin{equation}\label{ec:mass}
\frac{\partial \rho_f}{\partial t} + \nabla \cdot (\rho_f \textbf{v}_f) = 0
\end{equation}

\begin{equation}\label{ec:momentum}
\rho_f \frac{D \textbf{v}_f}{D t} = \nabla \cdot  \textbf{T}_f + \rho_f \textbf{b}_f + C_2 f(\phi) (\textbf{v}_f - \textbf{v}_T)
\end{equation}

where $\textbf{T}_f$ is the stress tensor of the fluid described as:

\begin{equation}\label{ec:stress tensor}
\textbf{T}_f = [-p(1-\phi)]\textbf{I} + 2\mu_f (1-\phi)\textbf{D}_f
\end{equation}

where $\mu_f$ is the asymptotic dynamic viscosity of blood. The scalar field $\phi$ is introduced to represent the volume fraction of deposited platelets (thrombus). The fluid phase $\rho_f$ is defined as

\begin{equation}\label{ec:density}
\rho_f = (1-\phi)\rho_{f0}
\end{equation}

where $\rho_{f0}$ is the fluid phase, $\textbf{b}_f$ is the body force, $\textbf{v}_f$ is and $ \textbf{v}_T $ are the velocity of the fluid and thrombus phases, respectively. $C_2$ is the resistance constant assuming that deposited platelets are densely compact spherical particles (2.78 $\mu$m), $f(\phi)$ is the hindrance function. For a full description of the original thrombosis model the reader is referred to the work of Wu et al.\cite{Wu2017} and Sorensen et al.\cite{Sorensen1999}.  

\subsection{Case of Study PFA-100\textregistered}
\begin{figure}[h!]
\centering
\includegraphics[width=0.28\textwidth]{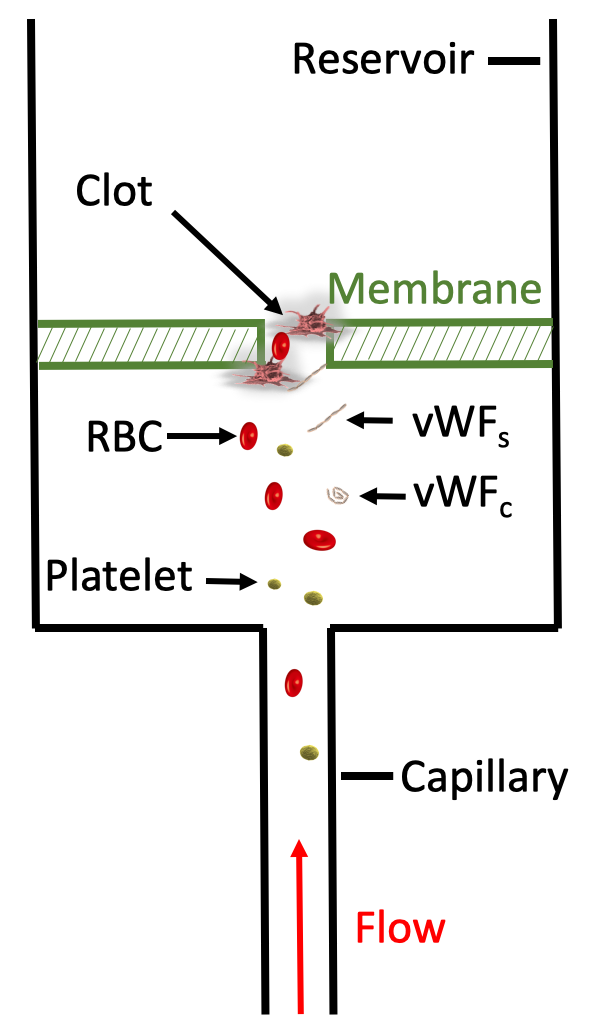}
\caption{\label{fig:PFA-100} Cross section of PFA-100\textregistered ~testing cartridge showing the capillary, the central membrane with orifice, and reservoir. As whole blood is aspired through the cartridge blood constituents aggregate in the coated membrane orifice.}
\end{figure}

The platelet function analyzer PFA-100\textregistered ~(Siemens Erlangen, Germany) is a coagulation testing device that is used to assess the primary hemostasis response \cite{Harrison2002}. Figure~\ref{fig:PFA-100} shows a cross section sketch of the PFA-100\textregistered ~test cartridge whose main component is a bio-active membrane with a central orifice of 140 microns diameter. Whole blood is aspirated through a capillary towards the membrane, as blood flows through the central orifice of the membrane, a thrombus forms until occlusion is achieved. The membrane is coated with collagen and epinephrine or ADP to promote thrombus formation. Flow is driven by a vacuum system that applies a $\Delta p$ of 4000 Pa. Shear rates inside the orifice reach values up to 6000 s$^{-1}$ promoting vWF unfolding leading to preferential thrombus formation\cite{Mehrabadi2016} inside the membrane orifice. When the orifice is fully occluded a Closure Time (CT) is obtained which is the quantity used to diagnose patients. The normal CTs are in the range 79-139 s for epinephrine cartridges and 61-105 s for ADP cartridges \cite{Harrison2002}. Prolonged CT times are observed in patients with major platelet defects (Bernard Soulier syndrome, Glanzmann's thrombasthenia, thrombocytopenia, etc.) and vWF disease,  or acquired vWF disease, among others \cite{Harrison2002}). In addition, CT times are affected by a variety of hematological and pharmacological factors such as hematocrit and aspirin.

\subsection{Baseline Thrombosis Simulation}\label{subsec:baselineThrombosis}
\begin{figure}[h!]
\centering
\includegraphics[width=0.65\textwidth]{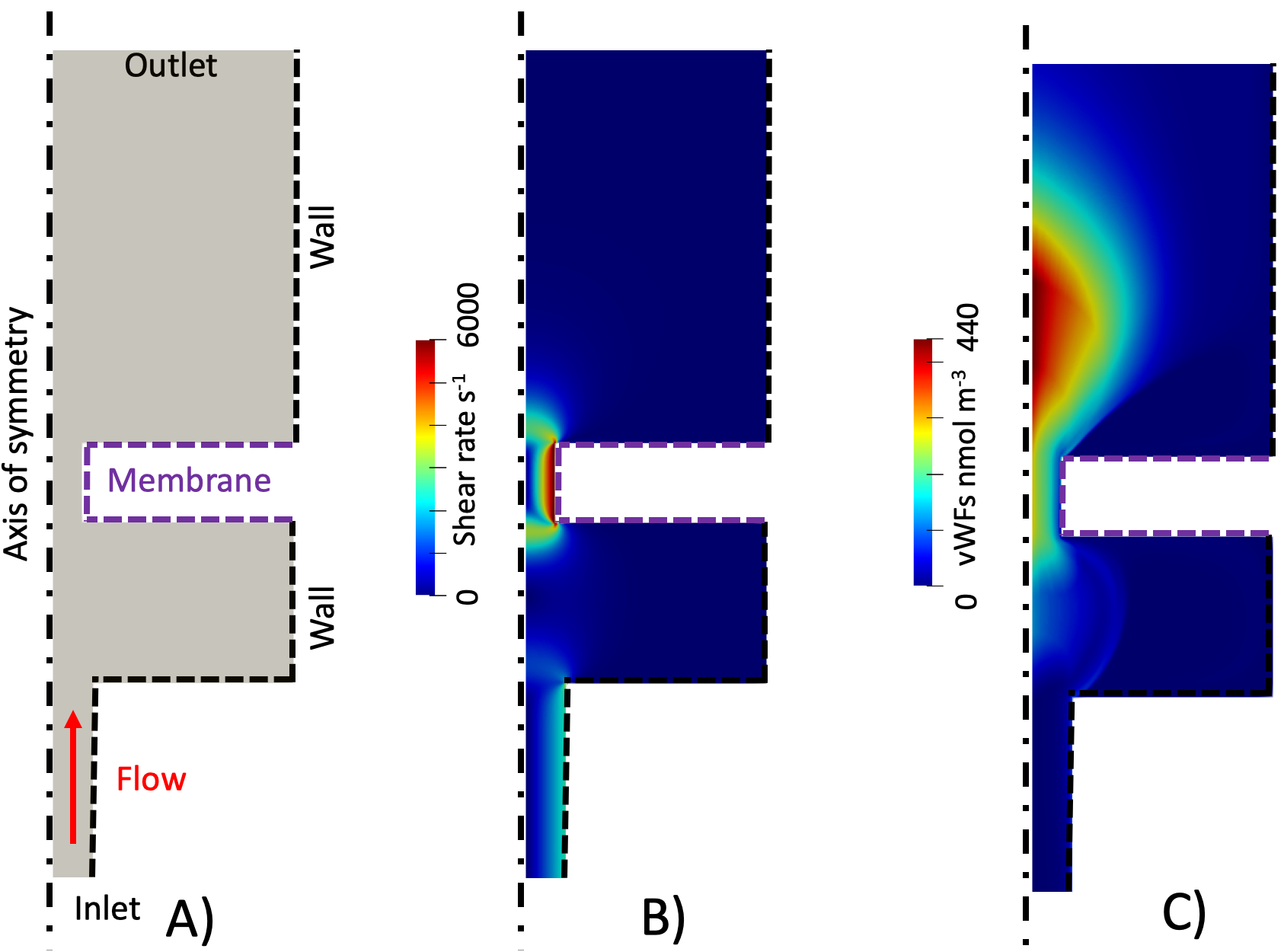}
\caption{\label{fig:PFA-SetUp} A) Numerical setup of PFA-100\textregistered ~thrombosis simulation. The membrane is set as reactive boundary condition allowing platelet deposition. B) Scalar shear rate field, large shear rates are observed in the membrane orifice. C) concentration of stretched vWF (vWFs) illustrating unfolding within and downstream of the membrane orifice.}
\end{figure}

An axisymmetric thrombosis simulation of the PFA-100\textregistered ~assay is the baseline case for the uncertainty quantification analysis. Figure~\ref{fig:PFA-SetUp}A shows the setup of the simulation. A uniform velocity profile $\textbf{v}_f = (0, 0.127, 0)$ m s$^{-1}$ was set at the inlet boundary. The membrane and cartridge walls were set as no-slip boundary conditions. A zero gradient velocity boundary condition was considered at the outlet. The pressure the outlet boundary was set to zero, and a zero gradient boundary condition was set for all remaining boundaries. In terms of biochemical species Table~\ref{tab:initial_conditions} lists the blood constituents concentrations that were prescribed at the inlet boundary. The platelet count values are representative of patients treated with ventricular assist devices \cite{Steinlechner2009}. At the outlet and cartridge walls a zero gradient boundary condition was applied for the species concentration. To reflect the ADP-collagen coated membrane, a reactive boundary condition was prescribed over the membrane where platelets can deposit. In addition, a diffusive flux of $ADP= 1 \times 10^{7}$ nmol m$^{-2}$ was prescribed to mimic shedding of ADP coating from the membrane. The ADP flux value was determined by best fit to clinical occlusion times reported by Steinlechner et al. \cite{Steinlechner2009}. All remaining variables were derived from the literature. The diffusion coefficient for each biochemical species were taken from \cite{Goodman2005}. We employed a Newtonian constitutive model for blood with a viscosity of 3.5 cP and a density of 1050 kg m$^{-3}$.

\begin{table}[h!]
\centering
\begin{tabular}{|c|c|}
\hline
Species & Concentration \\
\hline
RP &  216 $\times 10^3 $ Plt $\mu L^{-1}$ \\
AP &  2.16 $\times 10^3 $ Plt $\mu L^{-1}$ \\
\vWFc & 1000 nmol m$^{-3}$ \\
PT  &   1.1$\times 10^6 $ nmol m$^{-3}$  \\
TB  & 0 nmol m$^{-3}$ \\
AT  &  2.844  $\times 10^6 $ nmol m$^{-3}$ \\
ADP &  0 nmol m$^{-3}$ \\
TxA$_2$ &  0 nmol m$^{-3}$ \\
\hline
\end{tabular}
\caption{\label{tab:initial_conditions} Blood constituents inlet concentrations, the platelet count is representative of patients treated with ventricular assist devices \cite{Steinlechner2009}.}
\end{table}

\begin{figure}[h!]
\centering
\includegraphics[width=0.95\textwidth]{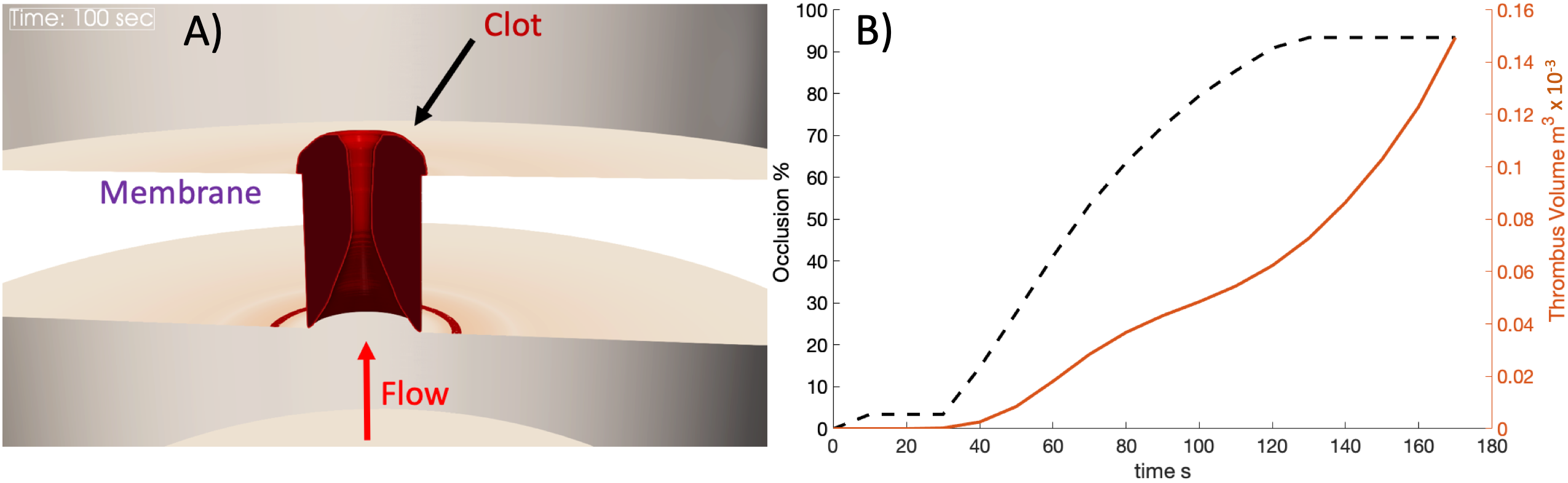}
\caption{\label{fig:PFA-Observables} A) 3D rotational extrusion view of thrombus formation at 100 seconds of simulation, clot forms inside the membrane orifice. B) Quantities of interest ~($Y(X_i)$) for the UQ analysis, occlusion percentage across the top edge of the membrane (dashed line) and total thrombus volume inside (solid line).}
\end{figure}

The mesh was composed of 68650 hexahedral cells with a finest resolution of  $\Delta x = 3$ $\mu$m located in the membrane orifice. A time step of  $\Delta t = 1 \times 10^{-3} $ s was used to solve the flow and species conservation equations with an open source finite-volume software package (OpenFoam). Before running the thrombosis simulation a steady state flow solution was obtained to improve stability at the first instants of the thrombosis simulation.
The resulting shear rate field was computed corroborating that inside of the membrane orifice induced shear rates greater than 6000 s$^{-1}$ required for vWF stretching. (See Figure~\ref{fig:PFA-SetUp}B and C.)

Figure~\ref{fig:PFA-Observables}A shows the thrombus formation at 100 seconds of simulation, the blood clot is visualized through a platelet volume fraction threshold > 0.8. Thrombus is formed inside the membrane orifice within the time scales observed clinically in VAD patients (80-150 s) \cite{Steinlechner2009}, the simulation closure time is defined at 80 \% of occlusion since full thrombotic closure is not numerically achievable due to exponential increase of mechanical shear prompting thrombus cleaning and preventing full occlusion. The main mechanism that drives clot formation can be explained as follows: 1) platelets deposit on the collagen coated membrane, 2) ADP within the membrane amplifies platelet activity, 3) large shear rates inside the membrane promote both platelet activation and vWF unfolding, further amplifying platelet deposition, 4) as the clot grows, the orifice area is reduced further increasing shear rates, which increases vWF activity forming a positive feedback loop mechanism. The quantities of interest (QoIs) chosen for the UQ analysis, described below, are: (1) the percentage of occlusion over the top edge plane of the membrane and (2) the total thrombus volume in the domain. These quantities are plotted in Figure~\ref{fig:PFA-Observables}B. 

\subsection{Uncertainty Quantification}
\subsubsection{Identification and modeling of random sources}
The thrombosis model used in this study is the most realistic we could assemble. In the following we will therefore put aside the model uncertainty in terms of its structure and underlying numerical algorithmic and we will focus only on parametric uncertainty. The model involves a total of 68 parameters (diffusion coefficients, reaction rates, biochemical concentrations, viscosity, density, etc). The uncertainties related to the microfluidic assay geometry were not considered as they are not as important as in physiological vessel hemodynamics. Due to the computational cost of performing an analysis considering all the model parameters, we have selected a subset of 15 parameters that based on our experience have a significant impact on the thrombosis formation dynamics. The parameters are listed in Table~\ref{tab:parameters} with their definition and distribution values. The literature does not show evidence of particular type of uncertainty distributions for these parameters. At best only low-order statistics are available (i.e. mean and standard deviation), which implies (under a maximum entropy principle) to model the distributions as normally distributed.  For sake of simplicity, independent Gaussian probability distributions were assumed for the parameters. Moreover, in order to introduce a similar level of uncertainty for all input parameters, we set the standard deviation to be ten percent of the mean value for each parameter, cf. Table~\ref{tab:parameters}. 
\begin{table}[h!]
\centering
\begin{tabular}{| l | c | c |}
\hline
Parameter $X_i$ & Definition & Distribution (mean, variance)\\
\hline
1) RP   & Resting platelets & $\mathcal{N} (216e3,21.6e3)$ Plt $\mu L^{-1}$ \\
2) AP & Activated platelets & $\mathcal{N}(2.16e3,0.216e3)$ Plt $\mu L^{-1}$ \\
3) \vWFc & vWF concentration & $\mathcal{N}(1e3,0.1e3)$ nmol m$^{-3}$ \\
4) $K_{rpd}$ & Deposition rate RP (to membrane) & $\mathcal{N} (1.0e\text{-}8, 1.0e\text{-}9)$ m s$^{-1}$ \\
5) $K_{apd}$ & Deposition rate AP (to membrane) & $\mathcal{N}(3.0e\text{-}6, 3e\text{-}7)$ m s$^{-1}$ \\
6) $K_{aa}$ & Deposition rate of AP to $AP_d$ & $\mathcal{N}(3.0e\text{-}5, 3e\text{-}6)$ m s$^{-1}$ \\
7) $K_{ra}$ & Deposition rate of RP to $AP_d$ & $\mathcal{N}(3.0e\text{-}6,3e\text{-}7)$ m s$^{-1}$ \\
8) $\dot{\gamma}_{vW\!F\!}$ & critical shear rate for vWF unfolding & $\mathcal{N}(5.5e3,0.55e3)$ s$^-1$ \\
9) $\Delta\dot{\gamma}$  & vWF shear unfolding transition width & $\mathcal{N}(1.271e3, 0.1271e3)$ s$^2$ \\
10) $\lambda_{crit}$ & Flowtype critical value for strong unfolding & $\mathcal{N}(5e\text{-}3, 5 e\text{-}4)$ dimensionless \\
11) $t_{vWF}$ & vWF relaxation time & $\mathcal{N}(0.05, 0.005)$ s \\
12) $Wi_{ef\!f\!,crit}$ & $Wi_{ef\!f\!}$ critical value for strong unfolding & $\mathcal{N}(0.316, 0.0316) $ dimensionless \\
13) $Wi_{ef\!f\!,hyst}$ & $Wi_{ef\!f\!}$ hysteresis value & $\mathcal{N}(0.053,0.0053)$ dimensionless \\
14) \vWFsCrit & Critical \vWFs concentration value & $\mathcal{N}(50, 5)$\\ 
15) $D^{h}_{e}/D^{l}_{e}$ & Ratio of adhesive forces at high and low shear rates & $\mathcal{N}(500,50)$ dimensionless \\
\hline
\end{tabular}
\caption{\label{tab:parameters}Fifteen (15) parameters of the thrombosis model.} 
\end{table}

\subsubsection{Polynomial chaos surrogate modeling framework}
In order to alleviate the computational cost of a parametric study involving the full thrombosis model, a surrogate model will be constructed in place of the direct thrombosis model. The surrogate model should approximate the quantity of interest $Y$ as accurately as possible.  It is constructed from a finite set of the thrombosis model predictions for selected parameter values in a "off-line" stage, and subsequently used "in-line" during the prediction stage. To this end, different methods are available such as Gaussian processes~\cite{Rasmussen2006}, support vector machines~\cite{Smola1998}, stochastic interpolation~\cite{Tatang97}, or stochastic spectral methods such as polynomial-based representations. 
Polynomial chaos expansions (PCE) will be retained to express the surrogate model in a closed form \cite{ghanem2003stochastic,le2010spectral,xiu2002wiener}. Let $(\Omega, \mathcal{B}, \mathcal{P})$ be the probability space where $\Omega$ is the space of random events $\omega$, this domain has a $\sigma$-algebra $\mathcal{B}$ and is equipped with a probability measure $\mathcal{P}$. The vector of normally-distributed random parameters can be written as ${\bX} \equiv {\bX}(\omega)=(X_{1}, \ldots, X_{d=15})$. If we consider, at a given time instant, the $d-$variate and second-order random variable $Y: \mathcal{I}_{{\bX}} \subseteq \mathbb{R}^d \rightarrow \mathbb{R}$, then $Y(t,{\bX}) \in \ell_2(\Omega, \mathcal{B}, \mathcal{P})$, can be expressed as the following expansion \cite{xiu2002wiener}:
\begin{align}
    Y(t,{\bX}) = \sum_{j=0}^{\infty} {{y}}_j(t)\, \mathrm{\psi_j}({\bX}),
    \label{gpc1}
\end{align}

where $\mathrm{\psi_j}({\bX}) = \prod_{i=1}^{d} \mathit{\psi_j^{(i)}}(X_{i})$ are the multivariate basis functions that form a complete basis, orthonormal with respect to the probability measure $\rho_{{\bX}}$ of the random input, and $\mathit{\psi_j^{(i)}}$ are the univariate Hermite polynomial functions along the $i^{\text{th}}$ dimension. 
% Moreover, we assume that all parameters are statistically independent.\\
In practice, PCE expansion must be truncated, and writes:\footnote{Instead of indexing the expansion on a single integer amounting to the cardinality of the entire approximation space, one can also make use of a multi-index notation that is equivalent. If $\Lambda_p$ is an index set for multi-index $\boldsymbol{\gamma} = (\gamma_1, \dots, \gamma_d) \in \mathbb{N}^d_0 $, then $\mathbb{P}_{\Lambda_p} \equiv \mathrm{span}\{{\psi}_{\boldsymbol{\gamma}} \  |\  \mathbf{\gamma} \in \Lambda_p\}$.
% and  we can then write $\mathrm{\psi_{\bf{\gamma}}}({\bX}) = \prod_{i = 1}^d \mathit{\psi_{\gamma_i}^{(i)}}(\xi^{(i)})$ where $\mathit{\psi_{\gamma_i}^{(i)}}$ is the $\gamma_i^{th}$ order basis function in dimension $(i)$.
}
\begin{align}
    Y(t,\boldsymbol{\xi}) \approx \sum_{\boldsymbol{\gamma}  \in \Lambda_p} {y}_{\boldsymbol{\gamma} }(t) \,\mathrm{\psi_{\boldsymbol{\gamma} }}({\bX}),
    \label{truncatedPCE}
\end{align}

where ${y}_{\boldsymbol{\gamma} }(t)$ are the evolving {\it modal} coefficients corresponding to the $\mathrm{\psi_{\boldsymbol{\gamma} }}$ basis.
We will restrict ourselves to tensor-product polynomial spaces $\mathbb{P}_{\Lambda_p}$, where $\Lambda_p$ is an index set of ``degree" $p$, and where $P = \mathrm{dim}({\mathbb{P}_{\Lambda_p}})\equiv \#\Lambda_p$, will denote the cardinality of the selected polynomial space. 
There are different ways of constructing the approximating polynomial spaces that will impact their cardinality. More specifically, in this study we will rely on $q-$norm type of polynomial spaces, allowing for an hyperbolic truncation of the approximation basis (therefore reducing the computational cost): 
         $\mathbb{P}_{\Lambda_{p,q}}$ with  index set
            $\Lambda_{p,q} = \{\by \in \mathbb{P}_{\Lambda_p} : \| \by \|_q = \left ( \sum_{\boldsymbol{\gamma}} y_{\boldsymbol{\gamma}}^q \right )^{1/q} \leq p \}$ \cite{Shen_SIAM2010}.   By default, the hyperbolic truncation with $q=1$ reduces to the standard total-degree truncation scheme. Decreasing the value of $q$ {\it de facto} decreases the number of polynomials of high interaction order kept in the expansion.

\subsubsection{Adaptively constructing the surrogate by least-squares minimization with regularization} \label{lsq}

Based on a training set $\{\bX^{(i)},Y^{(i)} \}_{i=1,\dots,N}$ collected at each time instant, ordinary linear regression can be used to compute the unknown coefficients $\boldsymbol{y} \equiv {y}_{\boldsymbol{\gamma}}$, \cite{choi2004polynomial,berveiller2006stochastic} by minimizing the residuals $\boldsymbol{r} \equiv \bY - \boldsymbol{\psi}_{\Lambda_p}
    \boldsymbol{y}$ in the $\ell_2-$norm through an optimization problem:
\begin{equation}
    \by = \argmin_{\boldsymbol{\mathrm{y}} \in \mathbb{R}^{P}} {\| \bY - \boldsymbol{\psi}_{\Lambda_p}
    \boldsymbol{\mathrm{y}}\|}_{2}^2,
    \label{regSol}
\end{equation}

where $\boldsymbol{\psi}_{\Lambda_p}$ is the measurement matrix corresponding to the gPC expansion in the index set $\Lambda_p$. The solution is obtained in matrix form as:
\begin{align}
    \by = \left( \boldsymbol{\Psi}_{\Lambda_p}^{\sf{T}} \boldsymbol{\Psi}_{\Lambda_p} \right)^{-1}\boldsymbol{\Psi}_{\Lambda_p}^{\sf{T}} \boldsymbol{\mathrm{Y}},
    \label{solLSQ}
\end{align}

where $\boldsymbol{\mathrm{Y}}$ is a vector of observations of size $N \times 1$, $\boldsymbol{\mathrm{\Psi}}_{\Lambda_p}$  the {\it measurement matrix} of size $N \times P$ with $\Psi_{ij}=\mathrm{\psi_j}({\bX_i})$, and $\boldsymbol{{y}}$ the vector of coefficients of size $P\times 1$. Unfortunately, when the number of simulations $N<P$, the problem is not well posed under this formulation and one needs to add a regularization term in the form of:
\begin{align}
    \boldsymbol{{y}} = \argmin_{\boldsymbol{\mathrm{y}} \in \mathbb{R}^{P}} \frac{1}{2} { \| \boldsymbol{Y} - \boldsymbol{\mathrm{\psi}}_{\Lambda_p} \boldsymbol{\mathrm{y}} \|}_{2}^2 + \lambda {\|\boldsymbol{\mathrm{y}}\|}_{1},
    \label{lassoDef}
\end{align}

with ${\|\boldsymbol{{y}}\|}_{1}=\sum_{{\boldsymbol{\gamma}  \in \Lambda_p}} \left | {y}_{\boldsymbol{\gamma}} \right |$, which is moreover a good strategy to favour sparsity of the surrogate in high dimension, i.e. that it forces the minimization to favour low-rank solutions.
In this paper, we rely on one of the several algorithms used to solve this constrained optimization problem, namely least angle regression (LAR) \cite{efron2004least}, which has been adapted to the context of PCE by Blatman and Sudret\cite{BLATMAN20112345}. They have designed an adaptive hybrid form of LAR in order to obtain a sparse PCE representation that remains reasonably accurate even with for small experimental designs. Other more robust sparse representation approaches favor the early detection of data outliers \cite{van2016robust}.
In practice the adaptive PCE-LAR algorithm will be run at each time instant at which the data sample from the full thrombosis model is available. For each surrogate build, it will test through a polynomial order range and will select the optimal polynomial based on some validation criteria described hereafter. Thanks to the formulation, the retained polynomial basis will most likely be quite sparse (which means that a large portion of the modal coefficients will be null), resulting in a compact surrogate model.
The leave-one-out (LOO) cross-validation error is a statistical learning technique, that is conveniently used for cases with small design of experiments (DoE). 
It consists in building $N$ metamodels, each one created on a reduced experimental design $\bX^l$, i.e. the training set obtained by removing a point $(\bX^{(l)},Y^{(l)})_{l \in \{1,2,\ldots,N\}}$, and comparing its prediction on the excluded scenario $\bX^{(l)}$ with the real value obtained from the thrombosis simulation. These $N$ errors are then averaged into a single value which may be collected for several time instants and different approximation bases for instance.

\subsubsection{Surrogate-based global sensitivity analysis} 
Once the modal coefficients are computed, moments, confidence intervals, sensitivity analysis and probability density function of the solution can be readily evaluated. Global variance-based sensitivity analysis may be performed to quantify  the relative importance of each (or a group of) random input parameter to the uncertainty response of the system.
The Sobol' functional decomposition of $Y=f(t,\bX)$ is unique and hierarchic. We have:
\begin{equation}
Y=f(t,\bX)=\sum_{\bs\, \subseteq \{1,2,\ldots N_d\}} f_{\bs} (t,\bX_{\bs}),
\label{eq:ANOVA}
\end{equation}
where ${\bs}$ is a set of integers such that $\bX_{\bs}=(X_{s_1},\ldots, X_{s_N})$, with $N=\textrm{card}({\bs})=|{\bs}|$ and $f_{\varnothing}=f_0=\mathbb{E}[Y]$. In this way, the variance of the response can be decomposed as \cite{ANOVA_Efron81}:
\begin{equation}
\mathbb{V}(Y) \equiv \sigma^2(t)=\sum_{{\bs}\, \subseteq \{1,2,\ldots N_d\}} \sigma_{\bs}^2(t), \quad \textrm{with} \quad \sigma_{\bs}^2(t) = \mathbb{V}\left ( \mathbb{E}\left [ Y | \bX_{\bs} \right ] \right )- \sum_{ \substack{\bu \subset {\bs} \\ \bu \neq {\bs} \\ \bu \neq \varnothing} } \sigma_u^2(t),
\end{equation}
where $\mathbb{V}$ and $\mathbb{E}$ are  the variance and the expectation operators, respectively.
The normalized Sobol' indices  $S_{\bs}$ are defined as \cite{Sobol_1993}:
\begin{equation}
\quad S_{\bs} \equiv  \frac{\sigma_{\bs}^2(t)}{\sigma^2(t)}  \quad \textrm{and}  \sum_{ \substack{  \bs\, \subseteq \{1,2,\ldots N_d\} \\ \bs \neq \varnothing}} S_{\bs}(t) = 1,
\label{eq:sobol_coeff1}
\end{equation}
which measure the sensitivity of the variance of $Y$ due to the interaction between the variables 
$\bX_{\bs}$, without taking into account the effect of the variables $\bX_{\bu}$  (for $\bu \subset \bs$ and $\bu \neq \bs$). 
In this work, we consider first- and total-order Sobol' indices. The first-order quantifies the effect of the single parameter alone $X_s$ on the output. The total index of input variable $X_s$, denoted $S^T_s$, is the sum of all the Sobol' indices involving this variable: $S^T_s=\sum _{\bs\, \subseteq \{1,2,\ldots N_d\}} S_{\bs}$.
This latter definition is not practical since it would result in computing each index separately. Instead, by denoting $S_{\tilde{s}}$ the sensitivity measure of all the variables {\it excluding} variable $X_s$, the total index can be rewritten as:  $S^T_s=1-S_{\tilde{s}}$.
Previous Sobol' indices can be evaluated by Monte-Carlo estimators, but require very large samples. Instead, they can be obtained very straightforwardly directly from the polynomial chaos modal coefficients \cite{UQdoc_14_104}.

\section{Results}
\subsection{Multi-constituent Thrombosis Simulations}
\begin{figure}[!h]
\centering
\includegraphics[width=1\textwidth]{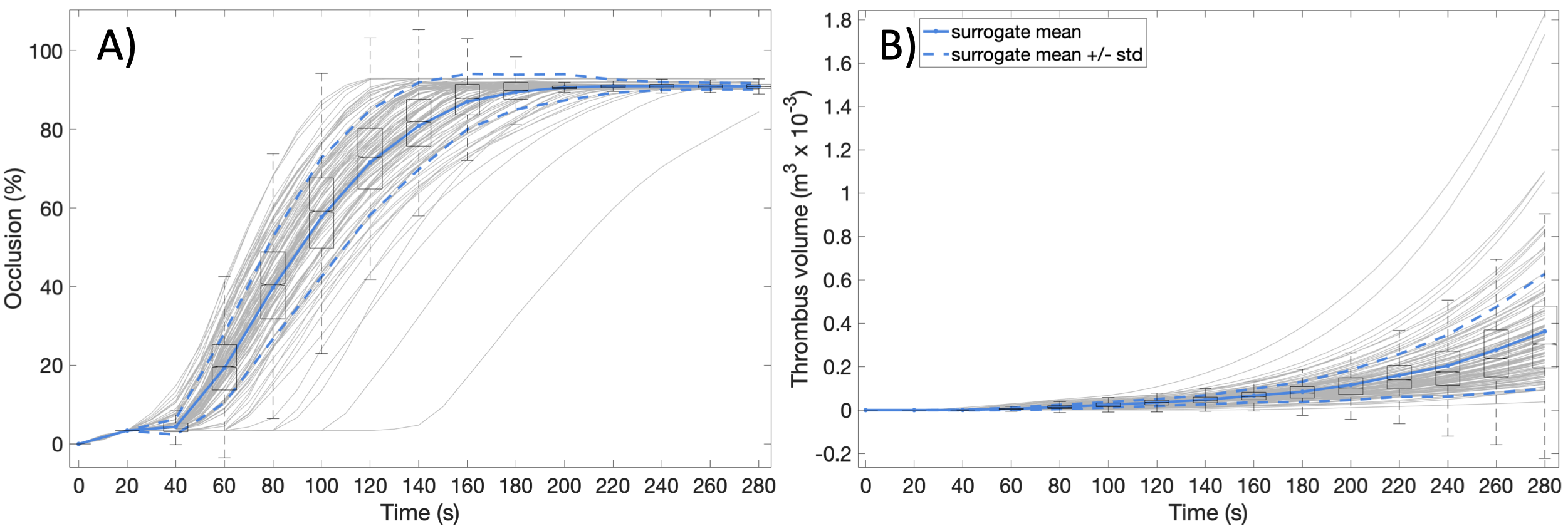}
\caption{\label{fig:Observables} Output variables for 120 thrombosis simulations (gray lines), the blue solid and dashed lines are the mean and standard deviation of the surrogate model respectively, the box plots show the median and 25$^{th}$ and 75$^{th}$ percentiles of the surrogate model predictions. A) Occlusion percentage evolution in time B) Thrombus volume in the computational domain in time.}
\end{figure}

The evolution in time of closure percentage and  thrombus volume is shown in Fig.\ref{fig:Observables} for 120 simulations. Closure percentage  (Fig.\ref{fig:Observables} A) is observed to follow a sigmoidal curve reaching a plateau of about 90\% at time of 100-200 seconds. The variance among simulations appears to be greatest during the rapid growth portion of the curve, and virtually nil during the first and last 25 seconds. It is noteworthy that the simulations do not reach full occlusion due to extremely high shear rates as the orifice cross section area approaches zero, and thrombus cleaning predominates. 

In terms of the thrombus volume, Fig.~\ref{fig:Observables}B shows little growth during the first 100 seconds, the curves start to deviate significantly around 125 seconds reaching different levels of total thrombus volume between approximately 0.1 to 1.7 $m^3 \times 10^{-3}$.  

Anticipating on the results of the following settings, both plots in Fig.~\ref{fig:Observables} also show the surrogate mean, std-based confidence interval, and a box plot of the 25$^{th}$ and 75$^{th}$ percentiles. It can be observed that the surrogate model is able to correctly reproduce the trend observed in the multi-constituent thrombosis simulations (gray plots).  

\subsection{PCE surrogate model training and validation}
\begin{figure}[h!]
\centering
\includegraphics[width=0.45\textwidth]{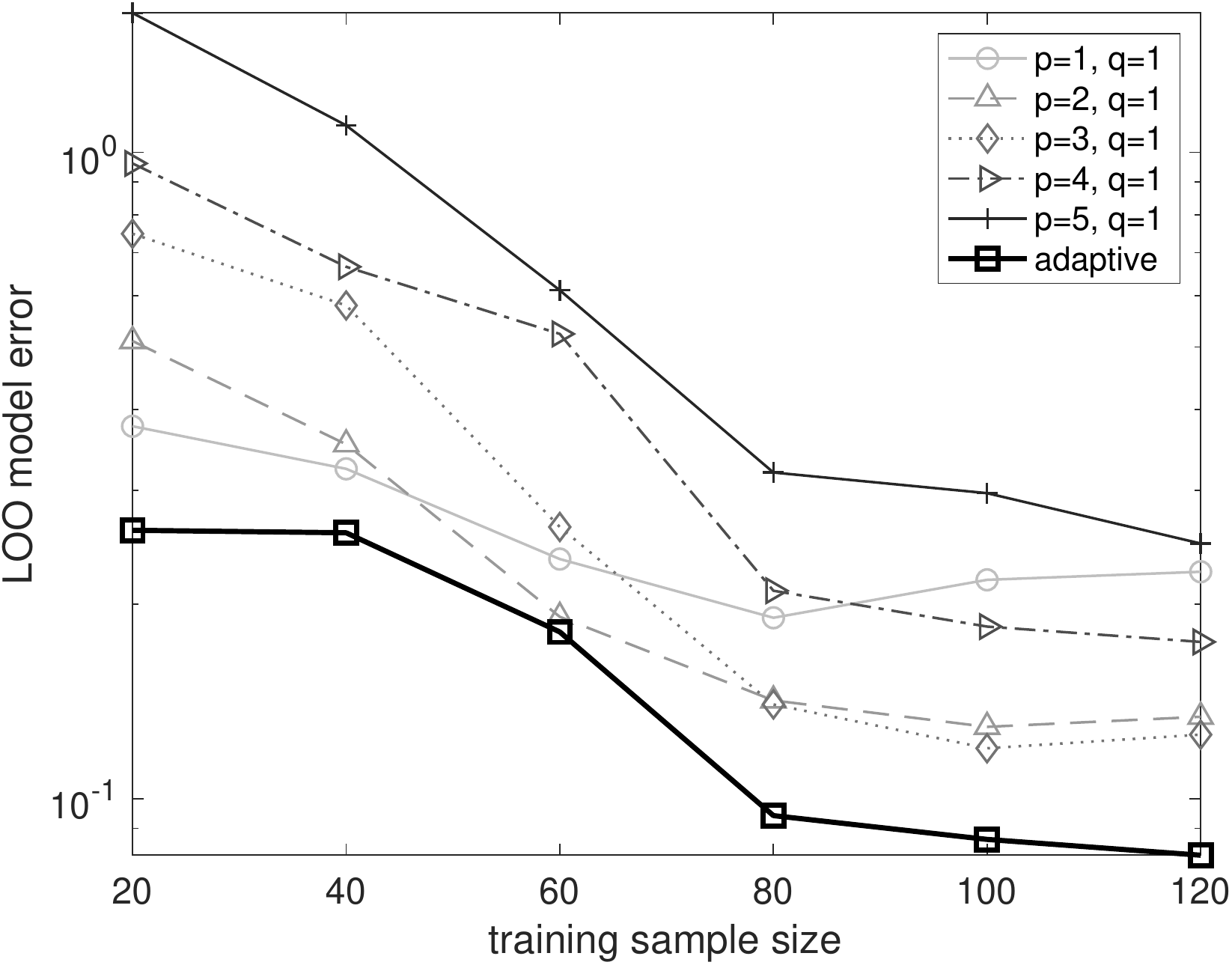}
\caption{\label{fig:LOOvsSample} Time-averaged leave-one-out cross-validation error (LOO) of the Thrombus Volume parametric model as a function of the training sample size and model complexity. Adaptive construction of the surrogate model is optimal with a reasonable budget of 120 full thrombosis simulations.}
\end{figure}

A preliminary study was performed to determine a reasonable number of simulations to build our training database, achieving a balance between predictive accuracy and computational cost. To ensure that the parameter space $\bX$ was uniformly explored, a Sobol' quasi Monte Carlo sequence was used to generate a DoE matrix. Then we computed the LOO error as a function of the training sample size using as QoI the closure time as define in Section~\ref{subsec:baselineThrombosis}. (See Figure \ref{fig:LOOvsSample}.) This was repeated for five surrogate models of increasing complexity (increasing polynomial order $p$ with truncation parameter, $q=1$). With exception of the linear model, the accuracy improved with increasing training size, reaching a point of diminishing returns at approximately 80-120 simulation. In terms of model complexity, the best performance was found for polynomial order from $p=2$ and 3. Further increasing the polynomial order was found to deteriorate accuracy, reflecting the limited sample size. This motivated the use of an adaptive approach that searches, for each time step, a combination of polynomial orders $p \in \{1,6\}$ and truncation parameters $q \in \{0.5:0.1:1\}$. This yielded lower error than fixed polynomial order and truncation parameter approximations. In conclusion, this pilot study  demonstrated that 120 simulations were sufficient when combined with an adaptive polynomial approach. 

\begin{figure}[ht]
\centering
\includegraphics[width=1\textwidth]{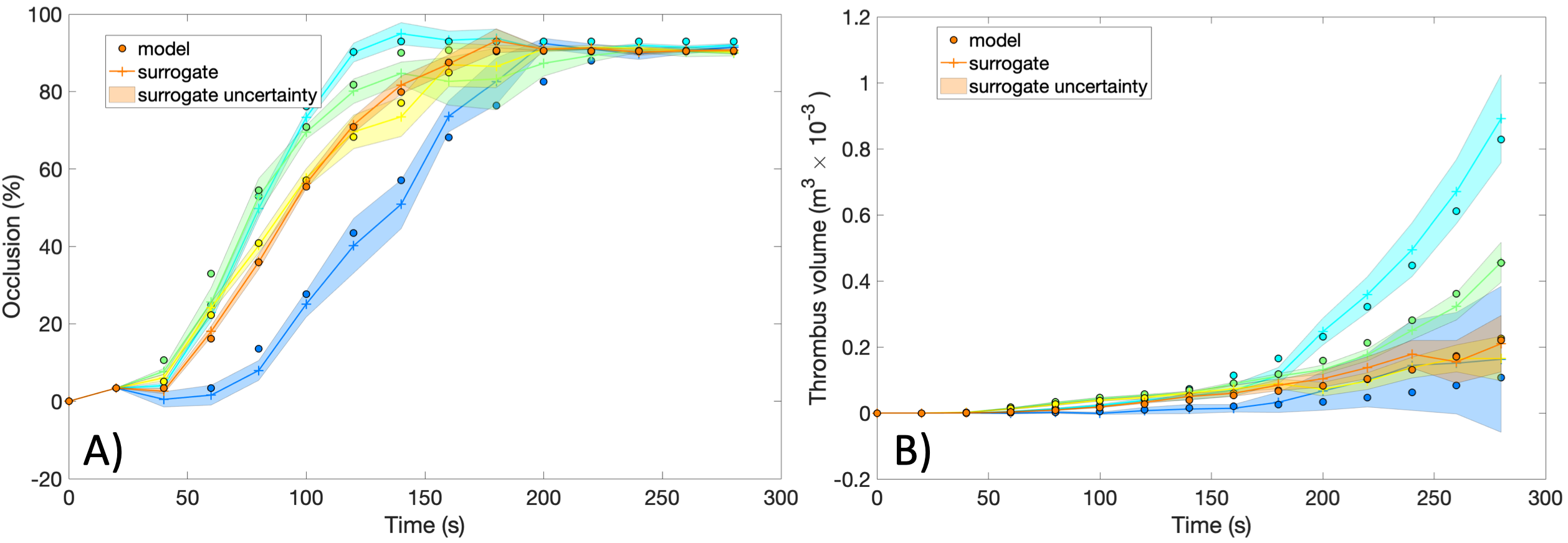}
\caption{\label{fig:thrombusVolumeCrossValidation} Cross validation: predictions from PCE surrogate model (solid line), thrombosis simulations (circles), the shaded area represents the confidence intervals associated with the meta-model (2 $\pm$ std).}
\end{figure}

When considering the closure percentage and the thrombus volume QoIs, a series of PCE meta-models were constructed, one for each time step at which the QoI was saved during the simulations. To cross-validate the model considering time evolution, 125 points were sampled using the quasi-random Sobol' sequence, of which 120 points were used for the training set to build the surrogate model and 5  points used as the test set. Five predictions using this methodology are shown in Figure~\ref{fig:thrombusVolumeCrossValidation}, the dots correspond to the thrombosis simulation, the solid line is the PCE surrogate and the shaded area is the PCE confidence interval computed using the bootstrap technique [Marelli and Sudret \cite{MARELLI201867}]. Producing a single prediction from each PCE surrogate is not reasonable due to the small training database and adaptive procedure used. Therefore, a confidence interval can be generated accordingly for each prediction. It is noteworthy that each surrogate model does not have the same selected polynomial approximation space (e.g. not same optimal order $p_{\text{opt}}$), depending on the complex nature of the response to the parameters randomness along time.

The accuracy of the PCE, evaluated by cross-validation with the LOO error criteria, remained for the occlusion percentage output under 20\% for most cases with only one time step at 30\% error (at $t=200$ seconds). In the case of the thrombus volume output the LOO error remained under 10\% reaching its maximum value by the last time step of the simulation. We consider that these errors showed that our surrogate was efficient considering that our initial space has $d=15$ dimensions and we only used a moderate computational budget of $N=120$ simulations. 

In conclusion, these results provided sufficient validation for our PCE approach to continue with the sensitivity analysis and further applications.

\subsection{Sobol' Indices}
\begin{figure}[h!]
\centering
\includegraphics[width=1\textwidth]{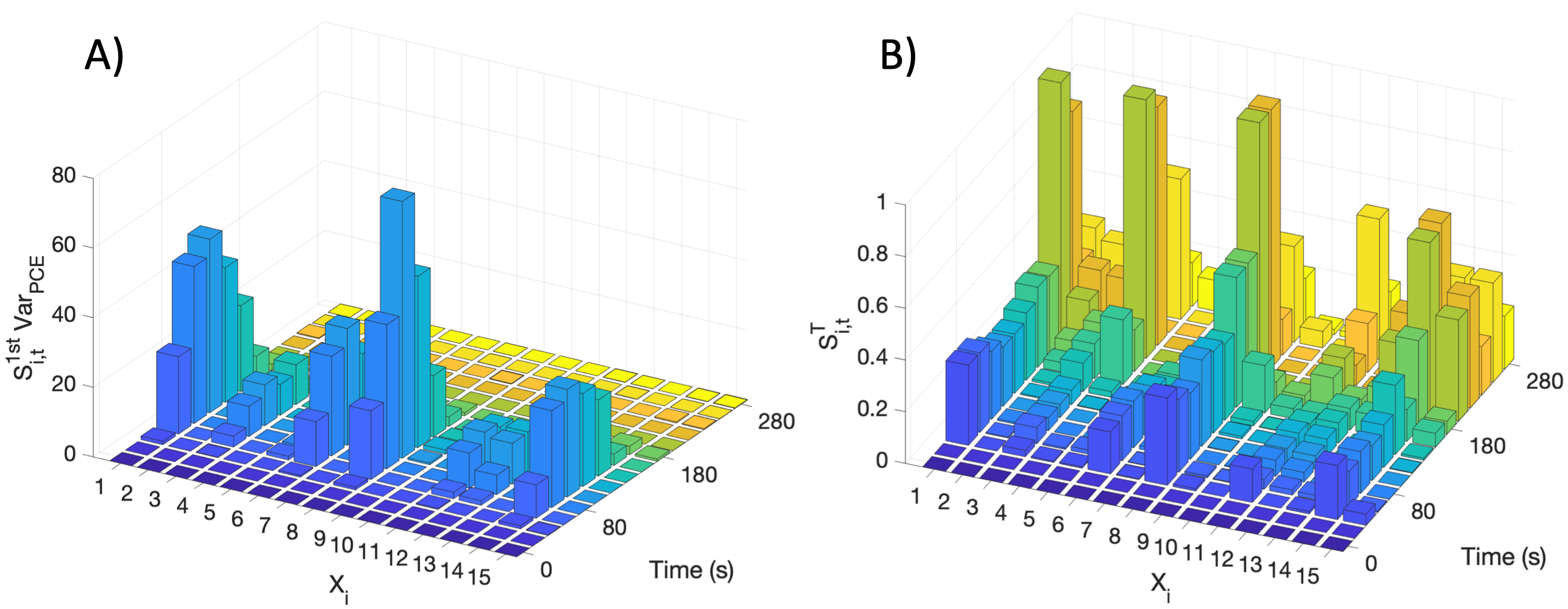}
\caption{\label{fig:Sobol1stOcclusion} Sobol' indices evolution over time for the QoI closure percentage. A) First order indices multiplied by the output PCE variance. B) Total Sobol' indices.}
\end{figure}

The first order and total Sobol' indices for the 15 parameters computed using the closure percentage as QoI are shown in Fig.~\ref{fig:Sobol1stOcclusion}. In order to scale the effect of each parameter, the first Sobol' indices were multiplied by the variance computed through the PCE approximation. Considering the first order indices, seven parameters have a significant impact in the range $50s<t<200s$ which is in line with the fact that most variations in the QoI are observed within this time range (see Fig.~\ref{fig:Observables} A). The physical significance of the parameters identified by the first Sobol' indices is related to three main factors: 
\begin{itemize}
\item \textbf{Thrombus constitution.} Resting platelet count (parameter 1) is the main constituent of the thrombus, therefore, it is understandably influential in the closure percentage.
\item \textbf{vWF activity.} The parameters 3, 8, 11, 12, and 14 ($vWF_c$, $\dot{\gamma}_{vWF}$, $t_{vWF}$, $Wi^{crit}_{eff}$, and $vWF^{crit}_{s}$) control the unfolding dynamics of vWF and its subsequent impact in platelet deposition. The fact that these parameters were flagged signals the important role of vWF in high shear rate thrombosis.
\item \textbf{Platelet-platelet adhesion.} The parameter 6 ($k_{aa}$) is the rate of platelet to platelet adhesion which directly influences the thrombus growth rate.
\end{itemize}

The Total Sobol' indices which evaluate the total effect of an input parameter, including all variance caused by its interactions with other parameters, are shown in Fig.~\ref{fig:Sobol1stOcclusion} B. Significant effects for 11 out of 15 parameters are observed indicating non-negligible parameter interactions. Interestingly large effects are observed specifically after 180 seconds even-though the QoI shows little variation after this time. This could be explained by the high order approximations that were retained for these time steps.

\begin{figure}[h!]
\centering
\includegraphics[width=1\textwidth]{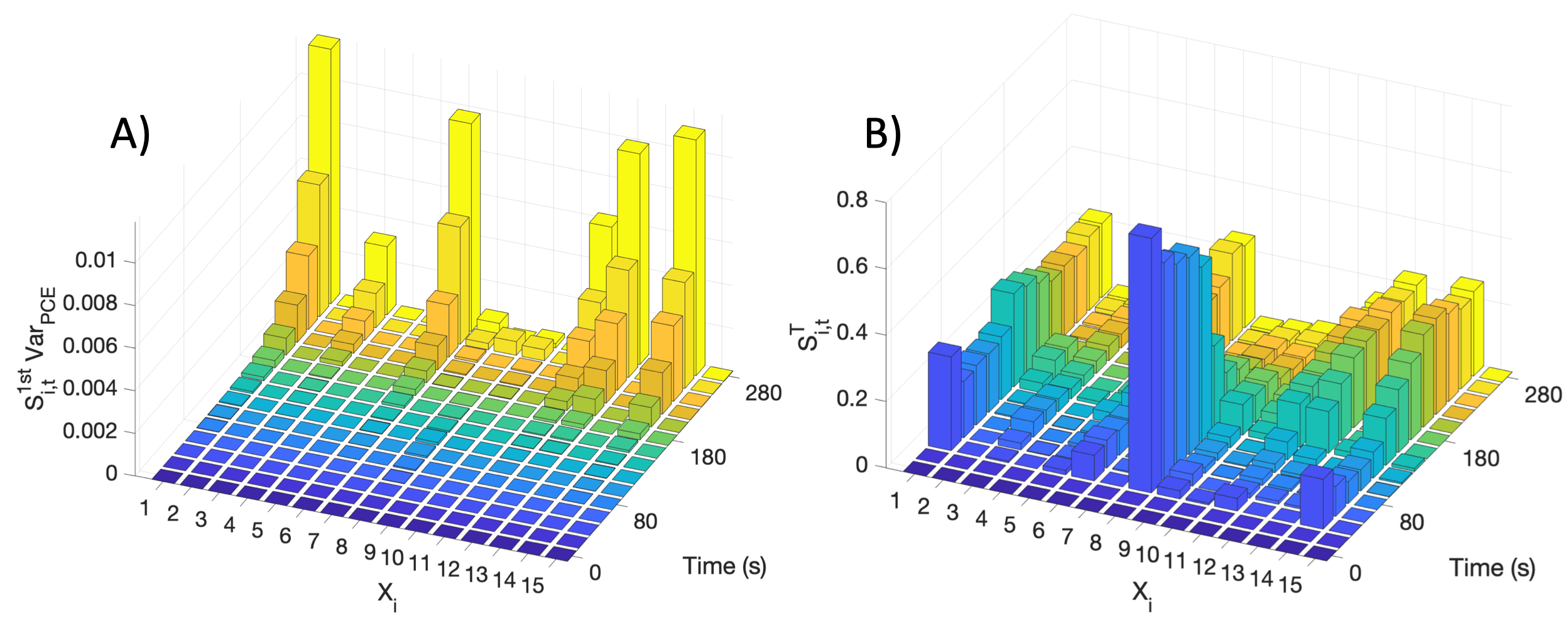}
\caption{\label{fig:ThrombusVolume} Sobol' indices evolution over time for the thrombus volume QoI. A) First order indices. B) shows the total Sobol' indices.}
\end{figure}

Figure~\ref{fig:ThrombusVolume} shows the first order indices multiplied by the PCE variance and the Total Sobol' indices for the total thrombus volume QoI. In this case, only six parameters showed significant contribution to the output variations mainly at the last time steps which is in agreement with the behavior observed in the time evolution of the total volume QoI (see Fig.~\ref{fig:Observables}B). As in the previous QoI, the  parameters that were identified are related to the thrombus constitution, the rate of platelet adhesion and vWF activity. These six parameters were also identified by the first Sobol' indices for the closure percentage QoI. Parameter 8 ($\dot{\gamma}_{vWF}$) that regulates vWF unfolding is not found to be influential in this case; although, a small contribution in the middle run of the thrombus formation is observed. This suggests that its role is only related to the initiation of the thrombus formation and not the subsequent growth of the thrombus. This is corroborated by the large value of its Total Sobol' index. (See Figure \ref{fig:ThrombusVolume}B.) The other six parameters found to be influential by the first Sobol' indices are also found to be influential considering their total Sobol' indices. Parameters 9 and 7 ($\Delta\dot{\gamma}$ and $k_{ra}$) show a low but non-negligible impact, these parameters are related to direct vWF activity and platelet deposition respectively. 

\subsection{Clinical Application: PFA-100\textregistered ~Closure Time}\label{sec:ResultsClinic}
To investigate the practical relevance and the generalization of the current UQ framework, the surrogate model predictions for three clinical scenarios were compared to reference clinical data: normal blood, thrombocytopenia, and vWF disease. The PCE surrogate model was calibrated using the closure time of the PFA-100\textregistered ~as QoI as defined in Section~\ref{subsec:baselineThrombosis}. The full set of thrombosis simulations (125) were used to compute the PCE coefficients. Three sets of 100,000 points (parameters) as defined in Table~\ref{tab:PCE_Clinical} were drawn using the Monte Carlo sampling technique described previously. The 13 supplementary model parameters that do not appear in Table~\ref{tab:PCE_Clinical} were held constant, as defined in Table~\ref{tab:parameters}. 

\begin{table}[h!]
\centering
\begin{tabular}{|c|c|c|}
\hline
\textbf{Case} & \textbf{Parameter} & \textbf{Distribution} \\
\hline
\textbf{Normal} & RP Resting platelet count & $\mathcal{N} (300e3,30e3)$ Plt $ \mu L^{-1}$  \\
\textbf{vWFD} & \vWFc (70 \% of vWF concentration) & $\mathcal{N}(700,70)$ nmol m$^{-3}$ \\
\textbf{Thrombocytopenia} & RP (50\% of Normal Resting platelet count)  & $\mathcal{N} (150e3,15.0e3)$ Plt $ \mu L^{-1}$ \\
\hline
\end{tabular}
\caption{\label{tab:PCE_Clinical} Definition of random parameters for the three cases.}
\end{table}

\begin{figure}[h!]
\centering
\includegraphics[width=0.7\textwidth]{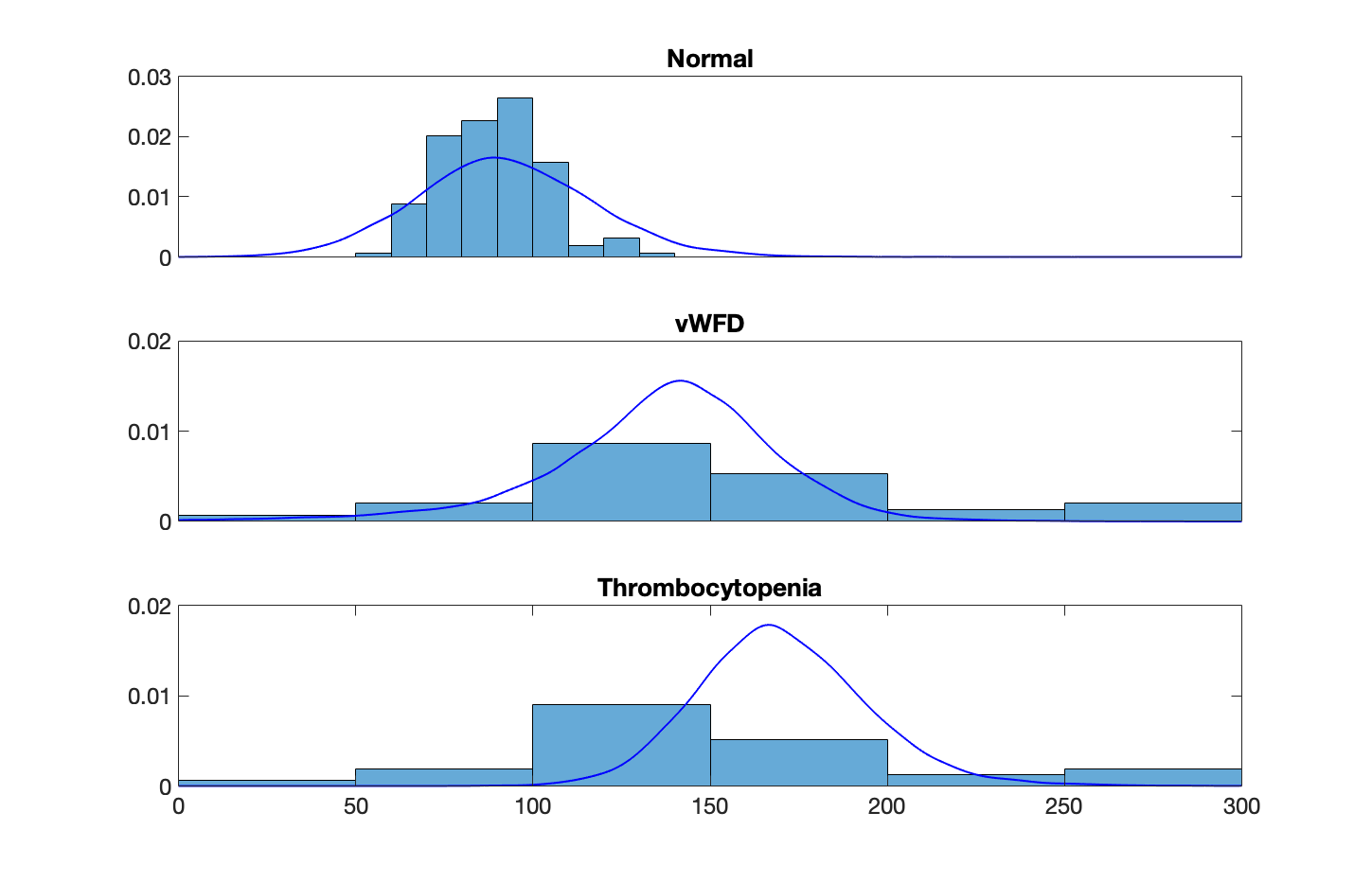}
\caption{\label{fig:PCEClossureTimes} Distribution of simulated and clinically reported PFA-100\textregistered ~closure time for normal blood, vWF disease and thrombocytopenia. The histogram corresponds to clinical data from \cite{Harrison2002} using ADP cartridges. The blue line is the probability density function computed with one hundred thousand realizations of the PCE surrogate model.}
\end{figure}

Once the input databases were created for each case. The surrogate model was used to predict the 100,000 closure times for each set of parameters. The probability distribution of the PCE closure times for each scenario is presented in Figure~\ref{fig:PCEClossureTimes} (solid lines) superimposed with clinical data derived from 77 patients, reported by Harrison et al. \cite{Harrison2002} demonstrating remarkable agreement; although the simulated closure time for vWFD and thrombocytopenia lag the clinical data slighltly. 

\section{Discussion}
Substantial progress has been made over the past decade in mathematical models of thrombosis which has enabled numerical simulation of a variety of clinically relevant thrombotic scenarios \cite{Peach:2014,WanAbNaim2018,Yazdani2021}. However, inherent uncertainties and assumptions of thrombosis models limit their translation to clinical practice and medical device development. This study investigated the impact of model parameter uncertainty in a thrombosis simulation by means of a polynomial chaos approximation. After extensive validation of the polynomial approximation, a global sensitivity analysis of 15 model parameters was performed. Six parameters showed significant influence in both quantities of interests considering their first Sobol' indices. This is in line with a recent study conducted by Melito et al. \cite{Melito2020} in which the authors found that only 4 of the 9 parameters in the model of Menichini \cite{Menichini2016} were influential for predicting volume fraction of thrombus in a backwards step. However, investigation of parameter interaction via the total Sobol' indices revealed influence of a subset of the 9 "non influential" parameters according to the first Sobol' indices. This suggest that simplification of the model must consider both first and total Sobol' indices, lest it neglects the interdependence of parameters.

Several limitations are acknowledged in the current study. First, the thrombosis model itself involves several approximations \cite{Wu2017}, in which blood is treated as a Newtonian fluid, and the roles of red blood cells and fibrin formation are not explicitly considered. Yet, in reality, the closure time is likely to be influenced by RBC trapping within the clot, and possibly influenced by platelet margination induced by RBCs. Second, the type of distribution and the magnitude of its variance that was assumed for the model parameters (10\%) was somewhat arbitrary and was applied to all the parameters equally. In future studies, each parameter should be assigned its own characteristic variance. In addition, the definition of closure time used in the simulations was based on 80\% occlusion of the orifice, whereas in reality it would be 100\%. This was necessary due to the simulations reaching an asyptotic value of approximatley 90\%  due to extremely high shear rates as the orifice diameter approaches zero. Finally, the clinical database for normal blood was constructed with n=50 patients; for the vWFD and thrombocytopenia, only n=15 patients and n=12 patients were available respectively. A larger sample size would improve the comparisons for vWFD and thrombocytopenia cases.

Despite its limitations, the PCE surrogate model demonstrated practical utility inasmuch as it permitted quantification of the uncertainty associated with a large number of parameters of a thrombosis model with economical computational cost. By contrast, it would have been very prohibitive to simulate all 300,000 sets of parameters with the full thrombosis model. Moreover, it was shown that the adaptive surrogate initially calibrated for certain parametric variabilities was nevertheless also capable of providing very reasonable predictions in parametric regions of lower probabilities. It is particularly noteworthy that despite the fact that the parametric distribution of the normal clinical case was centered outside the training confidence interval (estimated for this normal distribution as $1.3$ times the RP mean training value = $280e3$ Plt $\mu$ L$^{-1}$), the surrogate model nevertheless predicts very well the clinical observations. 

In conclusion this study showed that a PCE surrogate model trained with high fidelity thrombosis simulation can reproduce the trends observed in clinical data. These results encourage future studies using this methodology to investigate the influence of anti-platelet agents to identify scenarios that could achieve a balance between the risk of thrombosis and bleeding.

\section{Acknowledgments}
This research was supported by NIH R01 HL089456 and NIH R01 HL086918.

%\subsection{Bibliography}
\bibliographystyle{unsrt}  
\bibliography{ref.bib}

\end{document}